\documentclass[aps,pre,twocolumn,epsfig,floats]{revtex4}
\usepackage{amsmath}
\usepackage{color}
\usepackage{physics}
\usepackage{graphicx}
\usepackage{amsfonts}
\usepackage{mathrsfs}
\usepackage{amsmath}
\usepackage{amssymb}
\usepackage{float}
\usepackage[utf8x]{inputenc}
\usepackage{times}
\usepackage{hyperref}
\hypersetup{
  colorlinks=true,
  citecolor=blue,
  linkcolor=blue,
  urlcolor=blue}

\begin{document} 

\title{Dynamical route to ergodicity and quantum scarring in kicked coupled top}
 
\author{Debabrata Mondal, Sudip Sinha, and Subhasis Sinha} 
\affiliation{Indian Institute of Science Education and Research-Kolkata, Mohanpur, Nadia-741246, India}
 
\date{\today}

\begin{abstract}
Unlike classical system, understanding ergodicity from phase space mixing remains unclear for interacting quantum systems due to the absence of phase space trajectories. By considering an interacting spin model known as kicked coupled top, we elucidate the manifestation of phase space dynamics on local ergodic behavior of its quantum counterpart and quantum scarring phenomena. A transition to chaos occurs by increasing the kicking strength, and in the mixed phase space, the islands of regular motions within the chaotic sea clearly exhibit deviation from ergodicity, which we quantify from entanglement entropy and survival probability. Interestingly, the reminiscence of unstable orbits and fixed points can be identified as {\it scars} in quantum states, exhibiting athermal behavior and violation of Berry's conjecture for ergodic states. We also discuss the detection of quantum scars by a newly developed method of `out-of-time-order correlators', which has experimental relevance.
\end{abstract}


\maketitle

\section{Introduction}
Ergodicity of quantum many body system is a complex phenomena which has attracted significant interest in the recent years due to the advancement in cold atom experiments to study the non-equilibrium dynamics of many interacting particles \cite{Polkovnikov}. Quantum ergodicity is one of the key ingredients for understanding thermalization of isolated quantum system, which is the foundation of statistical mechanics, although not been fully understood. Unlike the case of classical ergodicity, which can be explained from chaotic dynamics leading to phase space mixing \cite{Zaslavsky,Eckmann,Casati}, the route to ergodicity in an isolated quantum system has been a long standing open problem. In this context, the eigenstate thermalization hypothesis (ETH) \cite{Deutsch,Srednicki} was proposed to explain ergodicity at the level of individual eigenstates, and its connection with random matrix theory (RMT) has also been explored in various quantum systems \cite{Izrailev_1,Izrailev_2,Kafri}. However, such mechanism does not provide a clear picture of the underlying phase space mixing leading to the ergodic behavior of closed quantum system in presence of interaction. Moreover, it is an important question to ask, how the non uniform mixing in mixed phase space region manifests in ergodicity of its quantum counterpart? The answer to this question is important for understanding the route to the deviation from ergodicity in an interacting quantum system. There are various examples of interacting quantum systems, which fail to thermalize and exhibit such deviation from ergodicity, the most popular being the systems showing many body localization (MBL) \cite{MBL1,MBL2,MBL3}. Besides MBL, the non ergodic phases have also been identified due to the presence of multifractal states \cite{Altshuler,Deng,Fazio_bhm}, which can also give rise to the anomalous thermalization \cite{Luitz}.
A recent experiment on a chain of Rydberg atoms \cite{Bernien} revealed absence of thermalization and periodic revival for some special initial state, which has been attributed to many body quantum scar (MBQS) \cite{Turner,Motrunich,M_Lukin1,abanin}. In addition, similar long lived nonthermal excited states have also been observed in a recent experiment on ultracold dipolar gas \cite{Lev}.  At the level of wavefunctions, the MBQS have also been identified in different interacting quantum systems \cite{spin1,onsager_scars,correlated_hopping,fracton,optical_lattice,K_saito,
scars_mps,sengupta_scars,AKLT,AKLT2,mondal,sinha1}. Originally, quantum scar has been identified as reminiscence of classically unstable trajectories in a non interacting system of chaotic billiards \cite{Heller}. However, the deviation from ergodicity due to the quantum scarring phenomenon and its connection with the underlying dynamical behavior in an interacting quantum system is not very clear and deserves more attention \cite{mondal,sinha1}.

In this work, by considering a periodically driven collective spin model, we explore the mixed phase dynamics and its manifestation in the ergodic behavior of its quantum counterpart. The kicked coupled top (KCT) model consists of two large spins interacting periodically among themselves, which has a suitable classical limit, allowing us to study the phase space dynamics. By increasing the kicking strength, KCT undergoes a smooth transition to chaos. Although, there are several methods to detect the quantum signature of chaos from eigenspectrum, they only provide an overall behavior without finer resolution due to the absence of its direct correspondence with classical phase space. In this context, it is a pertinent issue to investigate the ergodic behavior of a quantum system corresponding to mixed phase space dynamics with coexistence of regular region and chaotic sea, rather than featureless deep chaotic region. In KCT model, using the spin coherent states, describing the phase space semiclassically, we probe the local ergodic behavior of its quantum counterpart from entanglement entropy and survival probability, revealing the dynamical route to deviation from ergodicity. It is expected that with increasing degree of chaos, the dynamics of two spins become more entangled yielding enhanced entanglement entropy. Another property of ergodic evolution is the loss of memory of initial state, which can be quantified from the survival probability. In ergodic regime, the quantum states resemble the properties of random states and both entanglement entropy and survival probability converge to ergodic limit, independent of system parameters. From these quantities, we detect the local ergodic behavior as well its deviation due to the formation of quantum scar as a reminiscence of unstable dynamics. To elucidate the dynamical route to quantum scarring in this model, we simplified the dynamics into two classes corresponding to well known kicked top (KT) model \cite{haake_original,haake_book}, and the instability generated from mixing between them can lead to the formation of scars. We also demonstrate detection of scars by using the method of `out-of-time-order correlator' (OTOC), which is a newly developed tool to diagnose chaos in quantum systems \cite{stanford2,maldacena,K_hashimoto,Rozenbaum,Garttner1,
butterfly_effect,Swingle1,Fazio_otoc,A_M_rey2,Santos_otoc,Garcia_mata,
sray1,lakshminarayan,sudip_otoc} and implemented experimentally \cite{otoc_exp_1,otoc_exp_2}. Such method has also been applied in the context of black hole thermalization \cite{stanford2,maldacena}, and information scrambling \cite{Swingle1}, connecting the interdisciplinary areas of research.

The rest of the paper is organized as follows. In Sec.\ \ref{MODEL}, we introduce the Hamiltonian of KCT and discuss the Floquet formalism for stroboscopic evolution of the corresponding operators. Next, in Sec.\ \ref{CLASSICAL_ANALYSIS}, we derive the classical map for large spin limit and analyze the model classically. The fixed points and their stability is analyzed in subsection \ref{FP_stability}. The classification of the dynamics on reduced phase space and their correspondence with effective KT model is discussed in subsection \ref{DYNAMICAL_CLASSES}. The onset of chaos in KCT, as well the manifestation of phase space dynamics on quantum ergodicity  are presented in Sec.\ \ref{CHAOS_AND_ERGODICITY}. We investigate the scarring phenomena in Sec.\ \ref{QUANTUM_SCARS}: by identifying the scarred eigenstates, discussed in details in subsection \ref{FP_scars}, and their detection using the newly developed technique of OTOC in subsection \ref{FOTOC_scars}. Finally, in Sec.\ \ref{CONCLUSION}, we summarize the results and discuss the possible experimental detection of scars. The detailed derivation of stroboscopic evolution of spin operators is presented in appendix \ref{derivation_dynamical_map}. The instability exponents of the unstable fixed points of both KCT and effective KT model are compared in appendix \ref{stability_analysis}. In appendix \ref{quantum_scar_KT}, the quantum scarring phenomena in the effective KT corresponding to a dynamical class of KCT is discussed in details.
\section{model}
\label{MODEL}
 The periodically kicked coupled top (KCT) model \cite{ballentine,KCT} is described by the following Hamiltonian,
\begin{subequations}
\begin{eqnarray}
\hat{\mathcal{H}}(t)&=&\hat{\mathcal{H}}_0+\hat{\mathcal{H}}_c(t)\\
\hat{\mathcal{H}}_0&=&-\hbar\, \omega_0\,(\hat{S}_{1x}+\hat{S}_{2x})\label{H0}\\
\hat{\mathcal{H}}_c(t)&=&-\hbar\,\frac{\mu}{S}\,\hat{S}_{1z}\,\hat{S}_{2z}\sum_{n=-\infty}^{\infty} \delta(t-nT) \label{interaction}
\end{eqnarray}
\label{Hamiltonian_original}
\end{subequations}
where $\hat{S}_{ia}$ ($a=x,y,z$) represents the components of the spin  operators corresponding to two large spins ($ i=1,2$) of equal magnitude $S$. The Hamiltonian $\hat{\mathcal{H}}_0$ in Eq.\eqref{H0} describes the precession of two non interacting spins around the $x$-axis with angular frequency $\omega_0$, while the periodic kicking term with kicking strength $\mu$, represented by $\hat{\mathcal{H}}_c(t)$  in Eq.\eqref{interaction} periodically generates a ferromagnetic interaction between them with time period $T$. In rest of the paper and in all the figures, we scale energy (time) by $\omega_0$ ($1/\omega_0$) and set $\hbar=1$, $T=1$.

The quantum dynamics of a periodically kicked system is governed by its Floquet operator, describing the unitary time evolution between two successive kicks. The Floquet operator $\hat{\mathcal{F}}$ can be constructed from the free evolution $\hat{U}_0=e^{-\imath\hat{\mathcal{H}}_0 T}$ governed by the time independent Hamiltonian $\hat{\mathcal{H}}_0$ within a time period $T$, followed by an unitary operator $\hat{U}_c = e^{\imath\frac{\mu}{S}\hat{S}_{1z}\hat{S}_{2z}}$ describing the instantaneous kicking. Therefore, the Floquet operator can be written as \cite{haake_book},
\begin{eqnarray}
\hat{\mathcal{F}} = \hat{U}_c\hat{U}_0=e^{\imath\frac{\mu}{S}\hat{S}_{1z}\hat{S}_{2z}}e^{\imath (\hat{S}_{1x}+\hat{S}_{2x})T} 
\label{Floquet_operator}
\end{eqnarray}
For a periodically driven quantum system, the stroboscopic time evolution of an initial state $\ket{\psi(0)}$ can be written in terms of the Floquet operator $\hat{\mathcal{F}}$ as,
\begin{eqnarray}
\ket{\psi(n)} = \hat{\mathcal{F}}^{n}\ket{\psi(0)}
\label{time_evolution_floquet}
\end{eqnarray}
where $\ket{\psi(n)}$ is the state after $n^{\rm th}$ kick at time $t=nT$. Using the Heisenberg picture, the stroboscopic time evolution of an operator $\hat{\mathcal{A}}$ can also be written in terms of $\hat{\mathcal{F}}$ as, 
\begin{eqnarray}
\hat{\mathcal{A}}^{(n+1)}=\hat{\mathcal{F}}^{\dagger n+1}\hat{\mathcal{A}}\hat{\mathcal{F}}^{n+1}=\hat{\mathcal{F}}^{\dagger}\hat{\mathcal{A}}^{(n)}\hat{\mathcal{F}}
\label{operator_evolution}
\end{eqnarray} 
where $\hat{\mathcal{A}}^{(n)}$ denotes the operator at time $t=nT$. Similarly, for the present system, we obtain the stroboscopic evolution the spin components (see Appendix \ref{derivation_dynamical_map} for derivation) by setting $\hat{\mathcal{A}}=\hat{S}_{ia}$, which can be written as,
\begin{eqnarray}
\left[\hat{S}^{(n+1)}_{1x,1y,1z},\hat{S}^{(n+1)}_{2x,2y,2z}\right]^\text{T}=\hat{\mathcal{R}}\left[\hat{S}^{(n)}_{1x,1y,1z},\hat{S}^{(n)}_{2x,2y,2z}\right]^\text{T}
\label{quantum_map}
\end{eqnarray}
where $\hat{S}_{ix,iy,iz}$ represents the array of corresponding spin operators $(\hat{S}_{ix},\hat{S}_{iy},\hat{S}_{iz})$ and {\small T} denotes the transpose of the vector. The matrix $\hat{\mathcal{R}}$ generating time evolution can be represented in the block diagonal form as, 
\begin{equation}
\hat{\mathcal{R}}=
\left(\begin{array}{cc}
\hat{\mathcal{R}}_1&0\\
0&\hat{\mathcal{R}}_2\\
\end{array}\right)
\label{block_diagonal}
\end{equation}
The different blocks $\hat{\mathcal{R}}_i$ corresponding to the two spins (for $i=1,2$) are given by,
\begin{equation}
\hat{\mathcal{R}}_i=
\left(\begin{array}{ccc}
\cos \hat{Q}_{\bar{i}}^{(n)}& \sin \hat{Q}_{\bar{i}}^{(n)}\cos T & \sin \hat{Q}_{\bar{i}}^{(n)}\sin T  \\
-\sin \hat{Q}_{\bar{i}}^{(n)} & \cos \hat{Q}_{\bar{i}}^{(n)}\cos T  & \cos \hat{Q}_{\bar{i}}^{(n)}\sin T  \\
0&-\sin T & \cos T\\
\end{array}\right)
\label{individual_blocks}
\end{equation}
where $\hat{Q}_{\bar{i}}^{(n)}=\bar{\mu}(\hat{S}^{(n)}_{\bar{i}z}\cos T -\hat{S}^{(n)}_{\bar{i}y}\sin T )$ with $\bar{i}\ne i$ and $\bar{\mu}=\mu/S$. In the large spin limit, using the above stroboscopic evolution, we can obtain the classical map for the spins, which we discuss in the next section.

\section{Classical Analysis}
\label{CLASSICAL_ANALYSIS}
The classical limit of the above model can be achieved for large spin of magnitude $S$, and the corresponding classical dynamics can be studied for the appropriately scaled variables $\hat{s}_{ia}=\hat{S}_{ia}/S$, which behave classically, since the commutator $[\hat{s}_{ia},\hat{s}_{jb}]=i\epsilon_{abc}\delta_{ij}\hat{s}_{ic}/S$ vanishes in the limit $S\rightarrow\infty$. Classically, the spin vectors can be written as $\vec{s_i}\equiv(s_{ix},s_{iy},s_{iz})= (\sin\theta_i\cos\phi_i,\sin\theta_i\sin\phi_i,\cos\theta_i)$, where $\theta$ and $\phi$ denote its orientation, alternatively which can also be represented by the canonically conjugate variables $\phi_i$ and $z_i=\cos{\theta_i}$. By using Eq.\eqref{quantum_map}, the stroboscopic time evolution of the corresponding classical spin variables can be written as a classical map,
\begin{eqnarray}
\left[s^{(n+1)}_{1x,1y,1z},s^{(n+1)}_{2x,2y,2z}\right]^\text{T}=\mathcal{R}\left[s^{(n)}_{1x,1y,1z},s^{(n)}_{2x,2y,2z}\right]^\text{T}
\label{classical_map}
\end{eqnarray}
where $\mathcal{R}$ is the same matrix defined in Eq.\eqref{block_diagonal} and Eq.\eqref{individual_blocks}, except the fact that the operators $\hat{S}_{ia}$ are now replaced by the classical variables $s_{ia}$ and the coupling $\bar{\mu}$ becomes $\mu$. As a result, the stroboscopic map in Eq.\eqref{classical_map} becomes independent of $S$ and the condition $s^2_{ix}+s^2_{iy}+s^2_{iz} = 1$ is preserved for both the spins $(i=1,2)$ in the stroboscopic evolution.
\subsection{Fixed points and their stability}
\label{FP_stability}
The overall dynamical behavior is captured by analyzing the fixed points (FPs) and their stability, which is also important for understanding the ergodic behavior of the present model. The FPs can be obtained from the condition, $s^{(n)}_{ia}=s_{ia}^*$ (for all $n$). By analyzing the classical map given in Eq.\eqref{classical_map}, two trivial FPs T$_\pm$ are obtained and are given by, 
\begin{eqnarray} 
\{s_{1x}^*,s_{1y}^*,s_{1z}^*,s_{2x}^*,s_{2y}^*,s_{2z}^*\}=\{\pm 1,0,0,\pm1,0,0\}
\end{eqnarray} 
which remain stable for small kicking strengths $\mu<\mu_b=2\tan(T/2)$, as shown in Fig.\ref{fig:1_classical_analysis}(a) and \ref{fig:1_classical_analysis}(d). At the critical kicking $\mu_b$, both the T$_\pm$ undergo a pitchfork bifurcation and eventually become unstable, each giving rise to two new stable non trivial FPs with $s^{*}_{ia}\neq0$, which are denoted by NT$^{\rm L(R)}_{\pm}$  (see Fig.\ref{fig:1_classical_analysis}(a) and \ref{fig:1_classical_analysis}(e)), where the superscripts $\rm L(R)$  represent two new branches after bifurcation. The non trivial FPs NT$^{\text{L}(\text{R})}_\pm$, and their corresponding spin components can be obtained from the following equations, 
\begin{subequations}
	\begin{eqnarray}
	&&s_{iz}^{*2} \left[1+\tan^2\left(\frac{T}{2}\right)\text{cosec}^2\left(\frac{\mu}{2}s_{\bar{i}z}^*\right)\right]=1
	\label{Self-consistent equation}\\
	&&s_{ix}^*=\tan\left(\frac{T}{2}\right)\text{cot}\left(\frac{\mu}{2}s_{\bar{i}z}^*\right)s_{iz}^*\\
	&&s_{iy}^*=-\tan\left(\frac{T}{2}\right)s_{iz}^*
	\end{eqnarray}
	\label{Non-trivial-FP}
\end{subequations}
The sign of $s^{*}_{iy}$, $s^{*}_{iz}$ differs for the two bifurcated branches L and R, whereas the sign of $s^{*}_{ix}$ differs for the $\pm$ branches (see Fig.\ref{fig:1_classical_analysis}(a)). As the kicking strength increases, the FPs NT$^{\text{L}(\text{R})}_\pm$ become unstable at $\mu_u$, after that a period-doubling bifurcation at $\mu_{\text{\tiny TC}}=\pi/\cos(T/2)>\mu_u$ occurs, which leads to the formation of 2-cycles denoted by TC$^{\text{L}(\text{R})}_{2\pm}$ (see Fig.\ref{fig:1_classical_analysis}(a) and \ref{fig:1_classical_analysis}(f)). A 2-cycle describes the periodic oscillation between two specific phase space points, which can be obtained from the condition  $s^{(n+2)}_{ia}=s^{(n)}_{ia}$ for large $n$. The spin configuration corresponding to the pair of points of the 2-cycles TC$^{\text{L}(\text{R})}_{2\pm}$ are given by,
\begin{subequations}
\begin{eqnarray}
&&s'_{1x}=-s''_{1x}=\sqrt{1-\left(\frac{\pi}{\mu}\right)^2\sec^2 \frac{T}{2}}\\ 
&&s'_{1y}=s''_{1y}=\pm\tan\left(\frac{T}{2}\right)\frac{\pi}{\mu}\\
&&s'_{1z}=s''_{1z}=\mp\frac{\pi}{\mu}
\end{eqnarray}
\label{TC2+}
\end{subequations}
where the upper(lower) signs (in Eq.\eqref{TC2+}(b,c)) represents the 2-cycles originated from L(R) branches of the non trivial FPs. The components of the other spin of the same 2-cycles TC$^{\text{L}(\text{R})}_{2\pm}$ can be obtained from the conditions $s_{2x}=s_{1x},\,s_{2y}=\pm s_{1y},\,s_{2z}=\pm s_{1z}$, where $\pm$ denotes the 2-cycles originated from ${\rm NT}_{\pm}$. It is important to note that, apart from kicking strength $\mu$, the structure of FPs, 2-cycles and their stability strongly depend on the driving period $T$. For example, from Eq.\eqref{TC2+} it is evident that, TC$_{2\pm}$ exist only for $T<\pi$. We point out that in the present work, we restrict our discussion only for $T=1$.

Further increasing the coupling $\mu$, another pair of 2-cycles denoted by ${\rm TC}_{1\pm}$ emerge from the FPs ${\rm T}_{\pm}$. The spin components corresponding to one of the two points of the 2-cycles ${\rm TC}_{1\pm}$ are given by, 
\begin{subequations}
	\begin{eqnarray}
	&&s_{iz}^{'2} \left[1+ \cot^2\left(\frac{T}{2}\right)\text{cosec}^2\left(\frac{\mu}{2}s_{\bar{i}z}'\right)\right]=1
	\label{Self-consistent equation_2cycle}\\
	&&s_{ix}'=-\cot\left(\frac{T}{2}\right)\cot\left(\frac{\mu}{2}s_{\bar{i}z}'\right)s_{iz}'\\
	&&s_{iy}'=\cot\left(\frac{T}{2}\right)s_{iz}'
	\end{eqnarray}
	\label{TC1-}
\end{subequations}
Another point of these 2-cycles can be obtained from the condition, $s''_{ix}=s'_{ix}$, $s''_{iy}=-s'_{iy}$, $s''_{iz}=-s'_{iz}$. It is important to note that, for KCT, these pair of 2-cycles remain always unstable and are not very important, however they have significance in the context of dynamical classes, which is discussed in the next subsection and in appendix \ref{quantum_scar_KT}.

Apart from these, there is another pair of FPs denoted by FP-$\pi$, since the relative angle between the two spins is $\pi$, and are given by, 
\begin{eqnarray}
\{s_{1x}^*,s_{1y}^*,s_{1z}^*,s_{2x}^*,s_{2y}^*,s_{2z}^*\}=\{\pm 1,0,0,\mp1,0,0\}
\end{eqnarray}
which remain unstable for all kicking strengths. The FP structure and their stability with increasing kicking strength are summarized in Fig.\ref{fig:1_classical_analysis}(a). The details of the stability analysis and the instability exponents of unstable FPs and 2-cycles are given in appendix \ref{stability_analysis}. Here we only focus on these FPs which capture the essential features of the phase space, however with increasing kicking strength, more structure in the FPs and periodic cycles can be formed within a narrow range of $\mu$, that are not relevant for the present analysis.

\subsection{Dynamical classes and effective kicked top model}
\label{DYNAMICAL_CLASSES}
Since the model consist of two identical spins, the Hamiltonian and the classical dynamics remains invariant under the exchange of spins $\vec{S}_1 \leftrightarrow \vec{S}_2$. As a consequence, in terms of the redefined variables, $s_{a\pm}=(s_{1a}\pm s_{2a})/2$ $(a=x,y,z)$ or equivalently, $z_{\pm}=(z_1\pm z_2)/2$ and $\phi_{\pm}=(\phi_1 \pm \phi_2)/2$, the dynamics can be categorised into two subclasses with reduced phase space, namely: {\bf class} I. for which \{$s_{x-}=0,s_{y+}=0,s_{z+}=0$\} and {\bf class} II. for which \{$s_{x-}=0,s_{y-}=0,s_{z-}=0$\} holds, or equivalently both of them can be written as $\{z_{\pm}=0,\phi_{\pm}=0\}$. Note that, for both the classes, $s_{1x}=s_{2x}$. It can be verified from the classical map in Eq.\eqref{classical_map}, either of the above conditions (class I or II) remain valid for all coupling constant and the dynamics of the remaining variables reduces to that of the well known kicked top (KT) model \cite{haake_original,haake_book}, with (anti)ferromagnetic interaction corresponding to class (I)II.

Here we derive the equations of motion (EOM) corresponding to dynamical class I and show its correspondence with the effective antiferromagnetic KT model. Using the full dynamical equations (given in Eq.\eqref{classical_map}) and the constraints of the dynamical class I, the EOM of  remaining variables ($s_{x+},s_{y-},s_{z-}$)  are given by,
\begin{eqnarray}
\left[s^{(n+1)}_{x+,y-,z-}\right]^{\rm T} = \mathcal{R}_{-} \left[s^{(n)}_{x+,y-,z-} \right]^{\rm T}
\label{antiferro_EOM}
\end{eqnarray}
where the time evolution matrix $\mathcal{R}_{-}$ can be written as follows,
\begin{equation}
\mathcal{R}_{-}=
\left(\begin{array}{ccc}
\cos Q_{-}^{(n)}& -\sin Q_{-}^{(n)}\cos T & -\sin Q_{-}^{(n)}\sin T  \\
\sin Q_{-}^{(n)} & \cos Q_{-}^{(n)}\cos T  & \cos Q_{-}^{(n)}\sin T  \\
0&-\sin T & \cos T\\
\end{array}\right)
\end{equation}
with $Q^{(n)}_-=\mu(s^{(n)}_{z-}\cos T-s^{(n)}_{y-}\sin T)$. The above EOM can be shown to be the same as that of an effective antiferromagnetic KT model, which is described by the Hamiltonian,
\begin{eqnarray}
\hat{\mathcal{H}}(t)&=& -\hat{S}_{x} +\frac{\mu}{2S}\,\hat{S}^2_{z}\sum_{n=-\infty}^{\infty}\delta (t-nT)
\label{Anti KT hamiltonian}
\end{eqnarray}
where  $(\hat{S}_{x},\hat{S}_{y},\hat{S}_{z})$ are the components of the spin operator, which reduces to the classical variables $s_a=\hat{S}_a/S\,\,(a=x,y,z)$ in the limit of $S\rightarrow \infty$. The EOM corresponding to dynamical class I given in Eq.\eqref{antiferro_EOM} can be obtained from the classical map of the KT model given in Eq.\eqref{Anti KT hamiltonian}, under the change of classical variables $(s_{x},s_{y},s_{z}) \rightarrow (s_{x+},s_{y-},s_{z-})$. In a similar manner, one can also show, for the dynamical class II, the correspondence $(s_{x+},s_{y+},s_{z+})\rightarrow(s_{x},s_{y},s_{z})$ yields the EOM of a ferromagnetic KT model, where $\mu$ flips its sign in Eq.\eqref{Anti KT hamiltonian}.

\begin{figure}
\centering
\includegraphics[height=10.5cm,width=8.8cm]{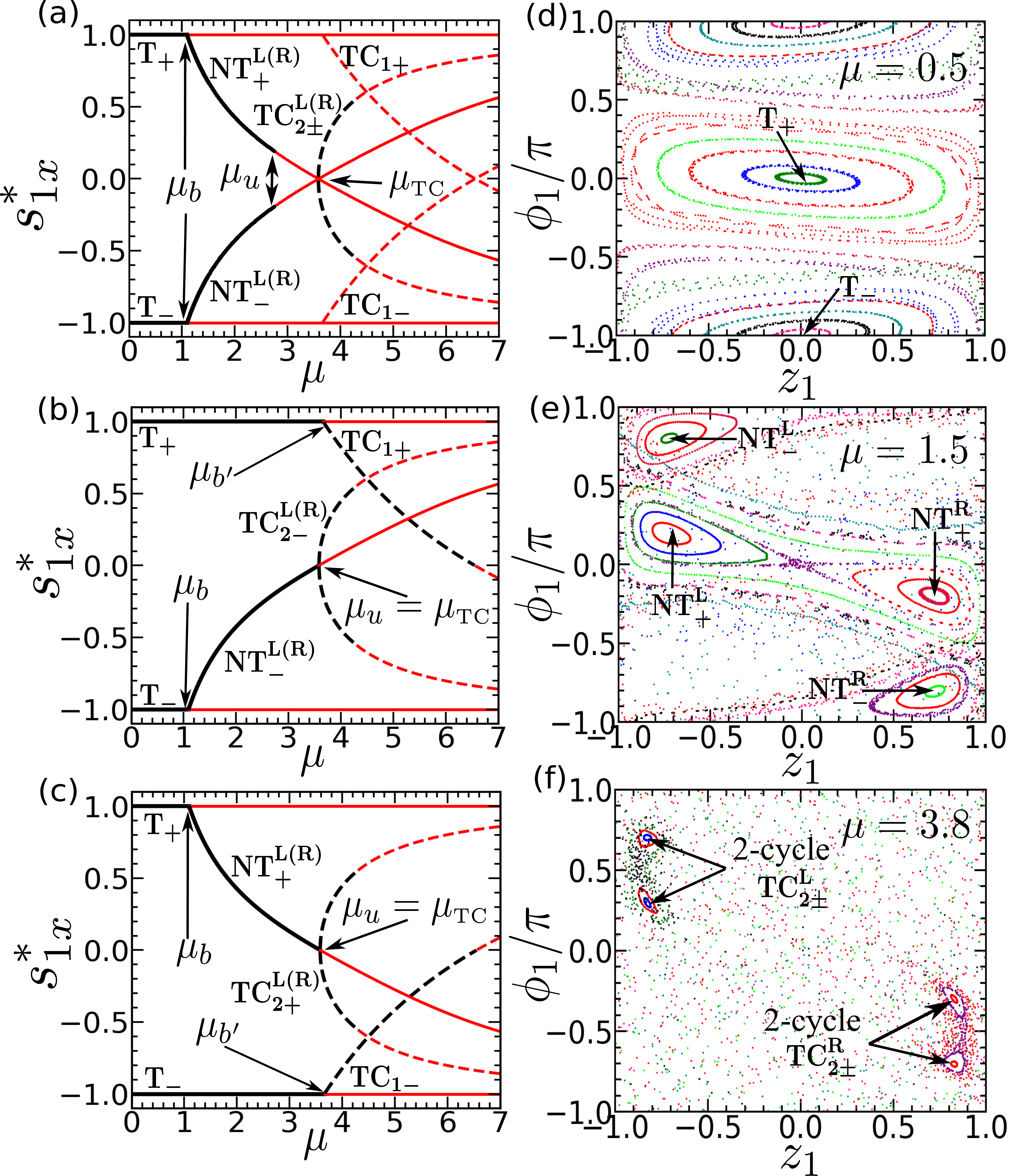}
\caption{Classical analysis of kicked coupled top (KCT): (a-c) Bifurcation diagram for different fixed points (FPs) and 2-cycles with increasing kicking strength $\mu$ for (a) KCT, which is compared with that of effective kicked top model with (b) antiferromagnetic (dynamical class I) and (c) ferromagnetic (class II) interaction. The stable (unstable) branches of FPs are denoted by solid black (red) lines, and the stable (unstable) branches of  2-cycles are denoted by dashed black (red) lines. The different  bifurcations and instabilities are marked by the arrowheads. (d-f) The phase portrait of KCT in $z_1$-$\phi_1$ plane for increasing $\mu$, pointing the various FPs and 2-cycles.}
\label{fig:1_classical_analysis}
\end{figure}

It is important to note that the actual phase space is not restricted by the constraints of the dynamical classes and the presence of initial perturbations violating the corresponding constraints leads to the mixing between the classes. Even when the dynamics is restricted to a particular dynamical class, the instabilities generated by the initially present small fluctuation can lead to the deviation from the corresponding class. As a result, the actual dynamics of the KCT model can deviate from that of the effective KT model. For clarification, we have shown the FPs and their stability for both the antiferromagnetic and ferromagnetic kicked top model (class I and II) in Fig.\ref{fig:1_classical_analysis}(b,c), and are compared  with that of the KCT model in Fig.\ref{fig:1_classical_analysis}(a). It is evident from Fig.\ref{fig:1_classical_analysis}(b,c), for dynamical classes I and II, the FPs and their stability exhibit complementary behavior. As a result, the FPs which are not present in the dynamical class I, such as NT$_+$, are present in class II. However, the FPs of both the effective KT models are present in the KCT model. Due to the presence of perturbations violating the constraints, the stable FPs and 2-cycles of a particular dynamical class become unstable in the KCT model for certain range of kicking strength, such as the unstable 2-cycles TC$_{1\pm}$ in KCT appear as stable 2-cycles in the corresponding KT model (see Fig.\ref{fig:1_classical_analysis}(b,c)). Such constraint violating fluctuations leading to the instability of the FPs plays a crucial role in the ergodic properties and formation of quantum scars, which is discussed in the later part of this work.

\section{Onset of chaos and ergodic behavior}
\label{CHAOS_AND_ERGODICITY}
\begin{figure}[H]
\centering
\includegraphics[height=10.2cm,width=8.7cm]{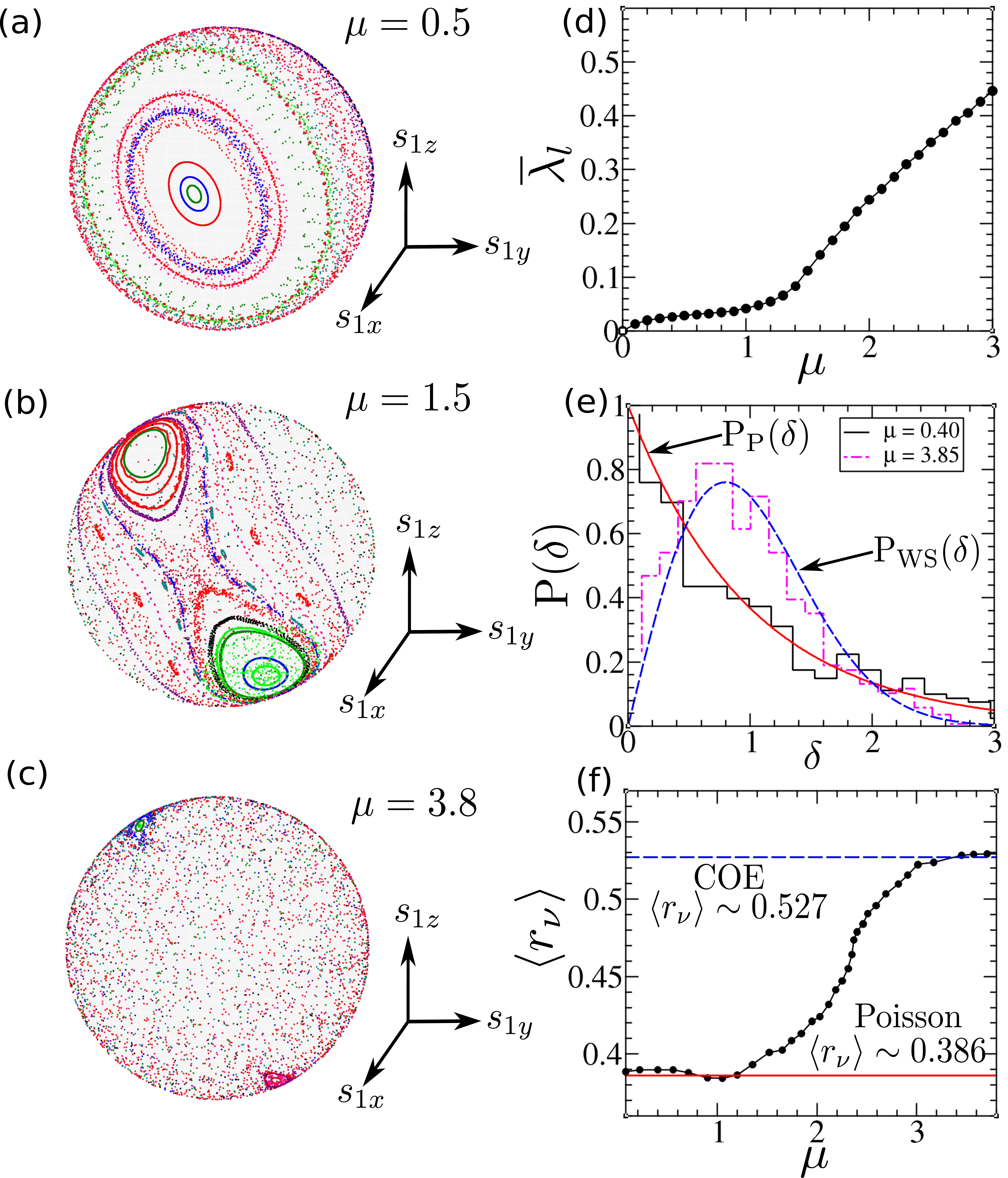}
\caption{(a-c) Phase portraits on Bloch sphere for increasing $\mu$, exhibiting onset of chaos. (d) Variation of average Lyapunov  exponent $\bar{\lambda}_l$ with increasing $\mu$. (e) Level spacing distribution of eigenphases of Floquet operator $\hat{\mathcal{F}}$ for two different values of $\mu$, exhibiting quantum signature of chaos. The solid red (dashed blue) lines denote the Poisson (Wigner-Dyson) statistics. (f) Average ratio of consecutive level spacings $\langle r_{\nu} \rangle$ with increasing $\mu$, showing a crossover from Poisson to Wigner-Dyson statistics. In this and all other figures, the quantum calculations are done for $S=20$ in KCT.}
\label{fig:2_onset_of_chaos}
\end{figure}

 After the bifurcation of trivial FPs at $\mu_b$,  more FP structures in the phase space appear, however the regular region around them shrinks and the trajectories in the remaining part become more irregular, as a result, a mixed phase space behavior is observed for intermediate kicking strengths (see Fig.\ref{fig:2_onset_of_chaos}(b)). Further increasing the kicking strength, the stable islands become unstable gradually, and the whole phase space is filled up with chaotic trajectories eventually, as shown in Fig.\ref{fig:2_onset_of_chaos}(c). Classically, the local chaotic behavior in the phase space can be identified by non vanishing Lyapunov exponent, which signals the exponential growth of initial perturbation with time \cite{lichtenberg,strogatz}. In the present analysis, the Lyapunov exponent $\lambda_l$ is numerically obtained by the method discussed in \cite{qr}. Since, the Lyapunov exponent in general can depend on the initial phase space point, to quantify the overall chaotic behavior, we compute the averaged Lyapunov exponent $\bar{\lambda}_l$ by averaging $\lambda_l$ over $\sim$ 4000 different initial phase space points. The onset of chaos in KCT is signalled from the sharp growth of $\bar{\lambda}_l$ with increasing kicking strength $\mu$ above $\mu_b$, as depicted in Fig.\ref{fig:2_onset_of_chaos}(d). The onset of chaos triggers the mixing in phase space, which is a key ingredient for classical ergodicity. 
 
Usually, the quantum signature of chaos can be detected from spectral statistics of the corresponding Hamiltonian. According to Berry Tabor's conjecture \cite{BT}, Poisson distribution of energy level spacing implies regular classical dynamics, whereas Bohigas-Giannoni-Schmit (BGS) conjecture \cite{BGS} suggests, Wigner-Dyson distribution of level spacing for a classically chaotic system. For periodically driven quantum systems, one can analyze the spectral statistics of eigenphases of the Floquet operator $\hat{\mathcal{F}}$. The eigenspectrum of $\hat{\mathcal{F}}$ is obtained from diagonalization, $\hat{\mathcal{F}}\ket{\phi_\nu}=e^{\imath \phi_{\nu}}\ket{\phi_{\nu}}$, where $\phi_{\nu}$ and $\ket{\phi_{\nu}}$ are eigenphases and corresponding eigenvectors of $\hat{\mathcal{F}}$, which contain relevant information related to the dynamics and ergodic properties. Numerically, the diagonalization of Floquet operator $\hat{\mathcal{F}}$ is done in the basis of $\hat{S}_{iz}$. 

In order to perform the spectral analysis corresponding to a particular symmetry sector, we identify two types of symmetries in the KCT model. The Hamiltonian in Eq.\eqref{Hamiltonian_original} remains invariant under the action of parity $\hat{\Pi} = e^{\imath \pi (\hat{S}_{1x}+\hat{S}_{2x})}$ and spin exchange ($S_{1}\leftrightarrow S_{2}$) operator  $\hat{\mathcal{O}}$ \cite{mondal}, which flips the indices of basis states $\ket{m_{1z},m_{2z}}$, where $m_{iz}$ are the quantum numbers of $\hat{S}_{iz}$. Both the operators posses two eigenvalues $\pm 1$, which we call as even (+1) and odd (-1). For spectral statistics, we only consider the eigenphases of the Floquet operator, for which the eigenvalue of $\hat{\Pi}(\hat{\mathcal{O}})$ are +1(+1). Next, the eigenphases are arranged within the range $[-\pi$,$\pi]$ in ascending order to compute the  corresponding level spacings, $\delta_{\nu}=\phi_{\nu+1}-\phi_{\nu}$. We calculate the normalized level spacing distribution keeping the mean to be unity, following the procedure as outlined in \cite{haake_book}.

\begin{figure*}
    \centering
    \includegraphics[height=16cm,width=17cm]{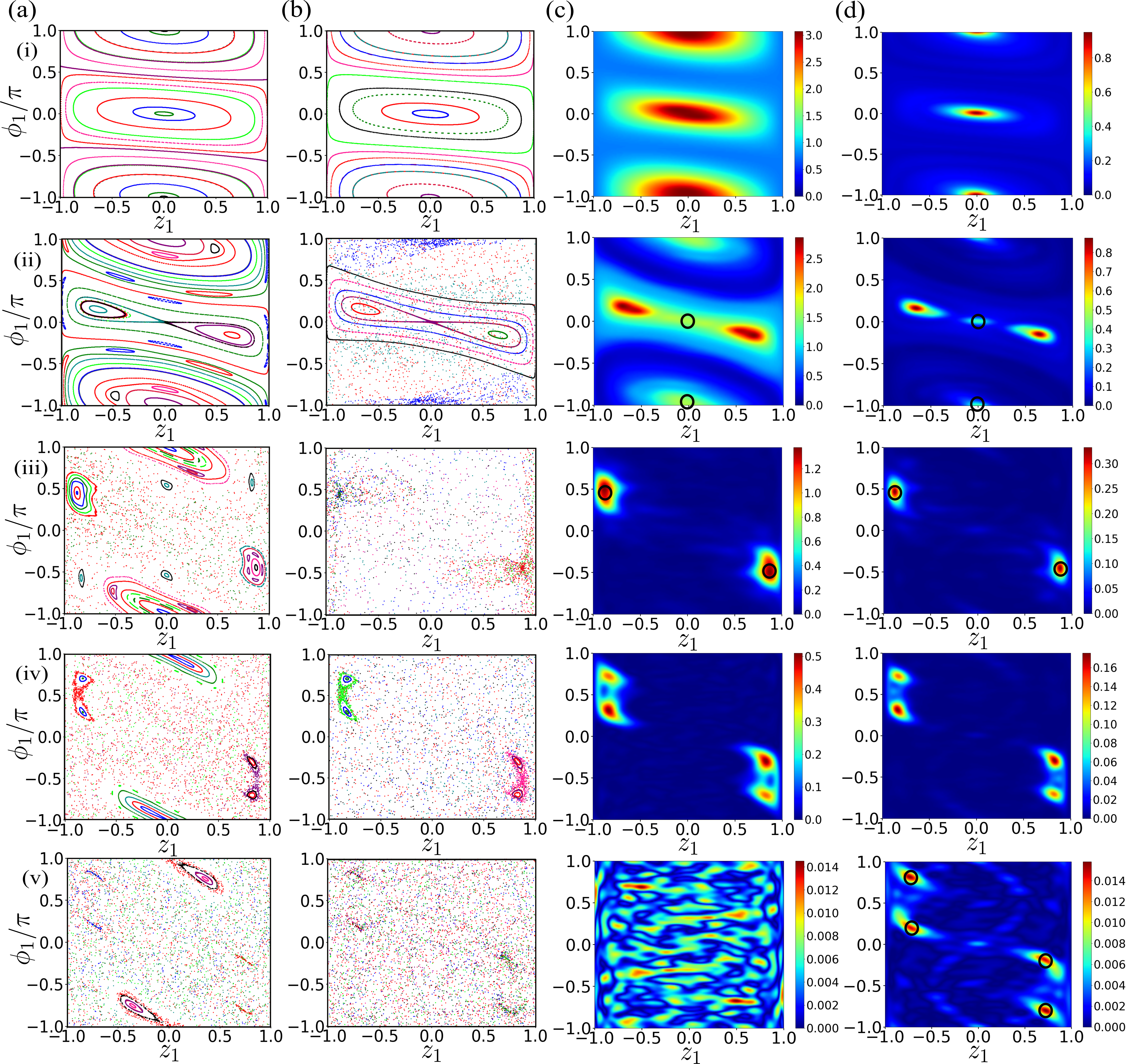}
    \caption{Reflection of phase space dynamics on local ergodic behavior of KCT quantified in terms of entanglement entropy and survival probability. Vertical columns: (a) Phase portraits for initial conditions belonging to dynamical class II (ferromagnetic KT). (b) Phase portraits in presence of small initial perturbations violating dynamical class II. (c) Color scaled plots of time averaged deviation of entanglement entropy  $\Delta S_{\rm en}$ and (d) survival probability $\Delta F$ from their ergodic limit. Initial coherent states correspond to dynamical class II (as in (a)). Due to the presence of intrinsic quantum fluctuations in coherent states, the constraint of dynamical class II is not maintained in quantum dynamics, thus the phase portraits in (b) are reflected on (c) and (d). The circles correspond to the unstable FPs in color scaled plots ((c) and (d)). The different rows correspond to  (i) $\mu = 0.5$, (ii) $\mu = 1.5$, (iii) $\mu = 3.22$, (iv)  $\mu = 3.8$, (v) $\mu = 4.34$. In the quantum dynamics, the time averaging is done from $n=50$ to $n=70$.}
    \label{fig:3_manifestation_of_phase_space}
\end{figure*}

As seen from Fig.\ref{fig:2_onset_of_chaos}(e), the resulting level spacing ($\delta$) distribution of eigenphases follows Poisson statistics, ${\rm P}_{\rm P}(\delta)=e^{-\delta}$ for smaller values of kicking strength, on the contrary, the spacing distribution shows level repulsion and approaches to Wigner-surmise, ${\rm P}_{\rm WS}(\delta) = (\pi\delta/2)e^{-\pi\delta^2/4}$ corresponding to orthogonal class of RMT for larger values of kicking strength above $\mu_b$, where the underlying phase space becomes fully chaotic. In addition, the average ratio of consecutive level spacings, $\langle r_{\nu} \rangle = \langle \rm min(\delta_{\nu},\delta_{\nu+1})/\rm max(\delta_{\nu},\delta_{\nu+1}) \rangle$ \cite{bogomonly_2} also exhibits crossover from Poisson statistics with $\langle r_{\nu} \rangle \sim 0.386$ to that of circular orthogonal ensemble (COE) of RMT with $\langle r_{\nu} \rangle \sim 0.527$ \cite{rigol_alessio} (see Fig.\ref{fig:2_onset_of_chaos}(f)).

Although, the spectral statistics reveals the underlying signature of chaos at the quantum level, the information about local chaotic behavior is still missing due to the absence of phase space description in quantum mechanics. To probe the local chaotic behavior, we use the prescription of spin coherent states \cite{radcliffe}, 
\begin{eqnarray}
\ket{\theta,\phi} = \left(\cos{\frac{\theta}{2}}\right)^{2S}\exp\left(\tan{\frac{\theta}{2}}e^{\imath \phi}\hat{S}_-\right)\ket{S_z = S}
\end{eqnarray}
with $\theta$ and $\phi$ representing the orientation of the spin vector $\vec{S}$, which provides a semiclassical description of phase space. To investigate the local degree of ergodicity, we evolve a coherent state, $\ket{\psi_c} \equiv \ket{\theta_1,\phi_1} \otimes \ket{\theta_2,\phi_{2}}$ corresponding to a particular phase space point for a sufficiently long time, and analyze the different properties of the final state $\ket{\psi(n)}$. First, we compute the reduced density matrix of the final state, $\hat{\rho}_{S} = {\rm Tr}_{\bar{\mathcal{S}}}\left(\ket{\psi(n)}\bra{\psi(n)}\right)$ by integrating out one of the spin sectors, which yields the entanglement entropy $S_{\rm en}$ as, 
\begin{eqnarray}
S_{\rm en} = -{\rm Tr}\hat{\rho}_{\mathcal{S}}{\rm log}\hat{\rho}_{\mathcal{S}}
\end{eqnarray}
It is expected that in the chaotic regime, the entanglement entropy increases with enhanced degree of chaos \cite{vidmar_rigol,Lewenstein,S_ghose1}, and in the extreme limit, it attains a maximum value $S_{\rm max}$ corresponding to a completely random state \cite{page}, which is given by,
\begin{eqnarray}
S_{\rm max} ={\rm log}(2S+1)-1/2
\end{eqnarray} 
Another characteristic feature of ergodic evolution is the loss of memory of the initial state, which can be quantified from survival probability. It is defined as the overlap of the time evolved state $\ket{\psi(n)}$ with the initial state $\ket{\psi(0)}$, 
\begin{eqnarray}
F(n) = \vert\langle \psi(n) \vert \psi(0)\rangle \vert^2 
\end{eqnarray} 
In the ergodic evolution, $F(n)$ decreases and at long time, saturates to limit $F_{\rm COE}=3/\mathcal{N}$ obtained from RMT \cite{Izrailev_2}, with Hilbert space dimension $\mathcal{N}=(2S+1)^2$. To probe the non ergodic behavior, we focus on the deviation of entanglement entropy $\Delta S_{\rm en} = \vert \bar{S}_{\rm en}-S_{\rm max} \vert$ and survival probability $\Delta F=\vert \bar{F}-F_{\rm COE} \vert$  from their ergodic limit. To eliminate the effect of temporal fluctuations, we obtain the time averaged value of the corresponding quantities denoted by $\bar{S}_{\rm en}$ and $\bar{F}$, where the time averaging is done over a certain interval towards the end of the stroboscopic evolution. For understanding the local ergodic behavior in phase space and to unveil its connection with the underlying classical dynamics, we compare $\Delta S_{\rm en}$  and $\Delta F$ with the corresponding classical phase portraits to investigate the dynamical route to local deviation from ergodicity. For clarity, here we only consider the dynamical class II (or equivalently the ferromagnetic KT model) defined by the constraint \{$z_{-}=0,\phi_{-}=0$\}, as well for quantum evolution, we choose the initial coherent states representing this class. We also investigate the changes in the phase portrait due to the presence of the small perturbations violating this constraint (see column 2 of Fig.\ref{fig:3_manifestation_of_phase_space}) and compare them with the phase portrait of dynamical class II (see column 1 of Fig.\ref{fig:3_manifestation_of_phase_space}). This comparison of classical phase portraits is important in the present context since such violation of constraints is inevitable in quantum evolution of coherent states of corresponding dynamical class, due to the presence of inherent quantum fluctuations. The manifestation of classical phase space dynamics on ergodic behavior of its quantum counterpart is evident from the comparison of $\Delta S_{\rm en}$ and $\Delta F$ with the phase portraits for different values of $\mu$, as depicted in Fig.\ref{fig:3_manifestation_of_phase_space}. The regular regions of phase space around the stable FPs leads to the strong deviation from ergodic behavior, which is quantified by enhancement of $\Delta S_{\rm en}$ and $\Delta F$, as seen in Fig.\ref{fig:3_manifestation_of_phase_space}(c,d). In the mixed phase space, the regular regions correspond to smaller $S_{\rm en}$ and larger $F$ values compared to the chaotic regions. It is also evident from Fig.\ref{fig:3_manifestation_of_phase_space}, the dynamics in presence of the constraint violating perturbations captures the ergodic behavior of its quantum counterpart more accurately.  Such local behavior of ergodicity quantified from $\Delta S_{\rm en}$ and $\Delta F$ elucidates its underlying connection with the corresponding dynamics.  
  
\section{Quantum scars}
\label{QUANTUM_SCARS}
In this section, we discuss the dynamical route to formation of quantum scars and their identification. From comparison of first two columns in Fig.\ref{fig:3_manifestation_of_phase_space}, it is clearly visible that, due to the presence of constraint violating perturbations of dynamical class II, certain FPs such as ${\rm T}_-$ become unstable and vanish from the phase portrait, however their reminiscence are still visible in $\Delta S_{\rm en}$ and $\Delta F$ as a {\it scar} of corresponding unstable FPs. As seen from Fig.\ref{fig:3_manifestation_of_phase_space}(d), $\Delta F$ is more capable of detecting scars compared to $\Delta S_{\rm en}$, as the chaotic region increases. It is important to note that, even when the FPs become unstable, the phase space trajectories still have a tendency to localize around them leading to the formation of quantum scars (see Fig.\ref{fig:5_NT_+_scar}(e,f)). The mixed phase space region gives rise to fascinating ergodic behavior, since stable FPs surrounded by the chaotic sea can coexist with the scars of unstable FPs, however the deviation from ergodicity is more prominent for stable FPs than that of quantum scars. 
\subsection{quantum scars of unstable fixed points and 2-cycles}
\label{FP_scars} 

\begin{figure}
\centering
\includegraphics[height=13.8cm,width=8.8cm]{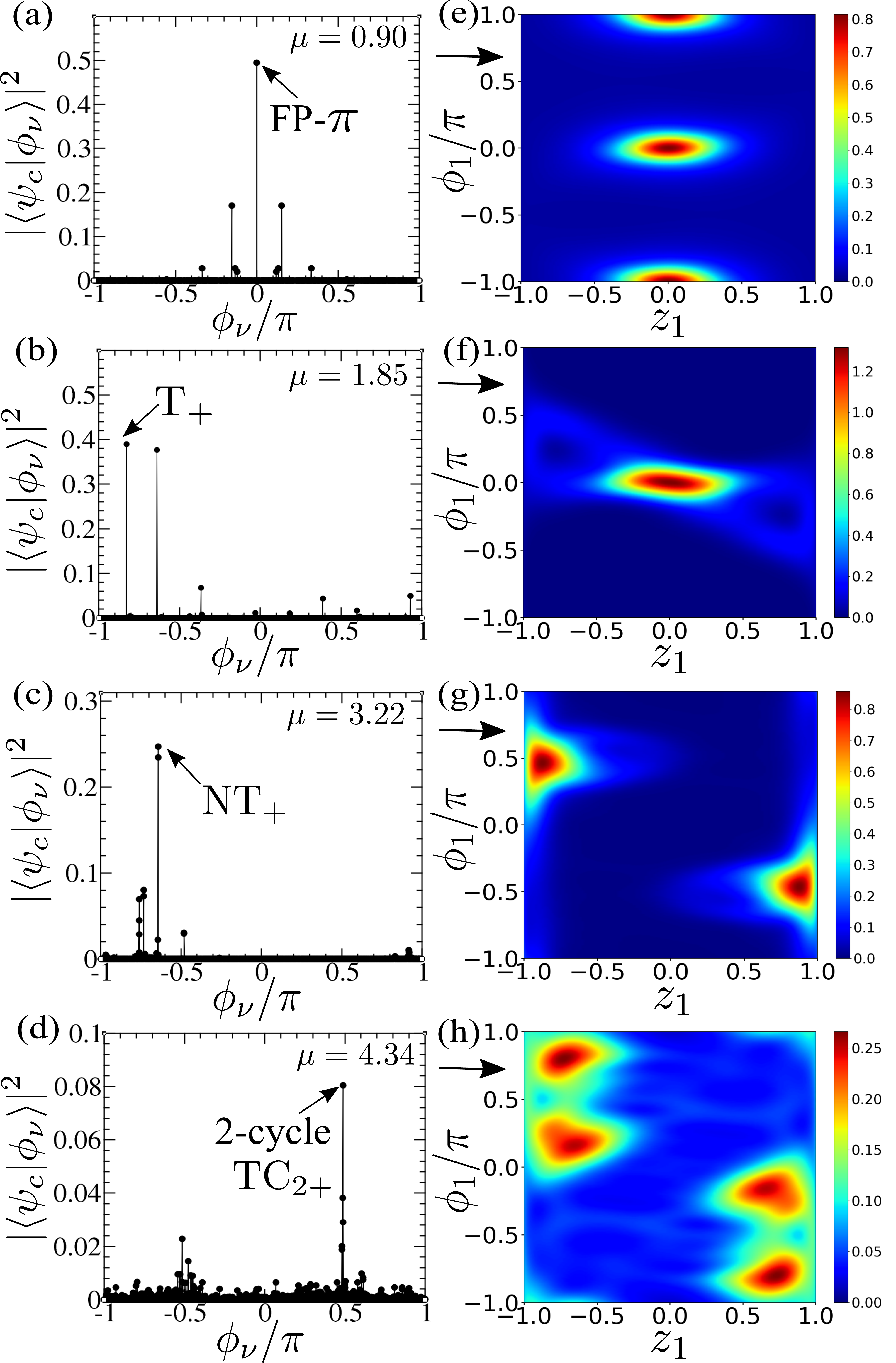}
\caption{Identification of different types of scars in Floquet eigenstates. (a-d) Overlap $|\langle \psi_c | \phi_{\nu}\rangle|^2$ of the coherent states $\ket{\psi_c}$ corresponding to different unstable FPs and 2-cycles (mentioned in the figure) with the Floquet eigenstates. The scarred eigenstates with maximum overlap are marked by the arrowheads. (e-h) Husimi distribution of the eigenstates having maximum overlap (as marked by the arrowheads in (a-d)) respectively, revealing the scar of the corresponding unstable FPs and 2-cycles.}
\label{fig:4_qs}
\end{figure}

\begin{figure}
\centering
\includegraphics[height=11cm,width=8.8cm]{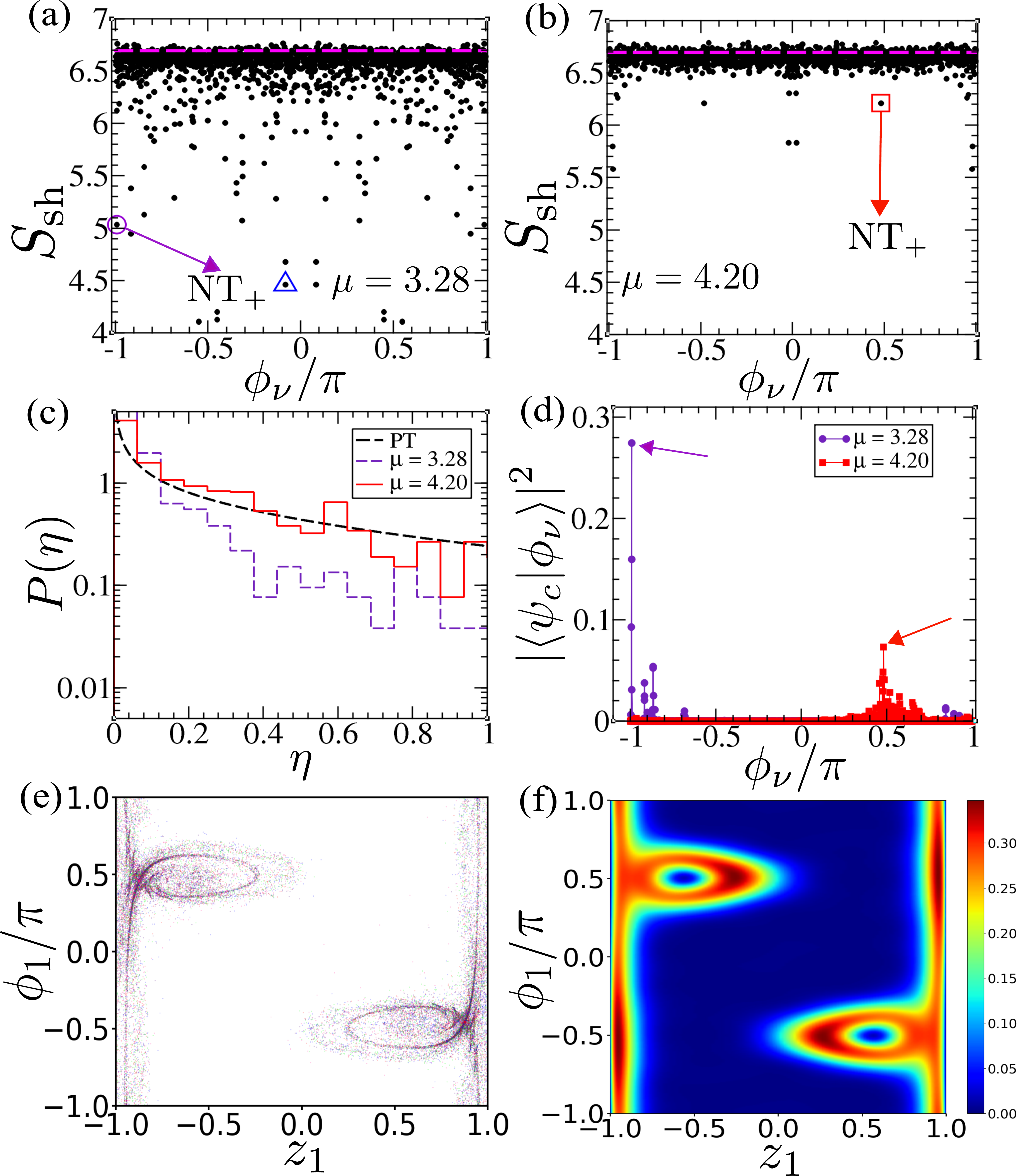}
\caption{(a-b) The Shannon entropy $S_{\rm Sh}$ of different Floquet eigenstates for two different values of $\mu$. The horizontal dashed line represents the COE limit and the eigenstates  containing the scar of NT$_+$ are marked by arrowheads. (c) The probability distribution $P(\eta)$ for scarred eigenstates of NT$_+$ (marked by arrows in (a-b)) for two different $\mu$ values. The black dashed line corresponds to the Porter-Thomas (PT) distribution. (d) The overlap $|\langle \psi_c | \phi_{\nu}\rangle|^2$ of coherent state corresponding to NT$_+$ with different Floquet eigenstates for different values of $\mu$ . The maximum overlap are marked by the arrowheads. (e) Unstable trajectories around NT$_+$ in presence of perturbations violating class II. (f) Husimi distribution of the deviated state (marked by triangle in (a)) depicting the scar of such unstable trajectory.}  
\label{fig:5_NT_+_scar}
\end{figure}

The scars of the unstable FPs, as identified from the deviation in entanglement entropy $\Delta S_{\rm en}$ and survival probability $\Delta F$ shown in Fig.\ref{fig:3_manifestation_of_phase_space}(c,d)), can also be detected in the Floquet eigenstates. The scarred eigenstates $\ket{\phi_{\nu}}$ can be identified from the large overlap $\vert \langle \psi_c \vert \phi_{\nu} \rangle \vert^2 \gg 1/\mathcal{N}$ \cite{abanin,sinha1,mondal} with the coherent state $\ket{\psi_c}$ representing semiclassically the unstable FP of corresponding scar (see Fig.\ref{fig:4_qs}(a-d)). On the contrary, it is expected that such overlap becomes $\sim 1/\mathcal{N}$ in the ergodic regime indicating complete delocalization. To visualise the scars, we compute the Husimi distribution of the reduced density matrix $\hat{\rho}^{\nu}_{\mathcal{S}}$ obtained from the scarred eigenstates $\ket{\phi_{\nu}}$,
\begin{eqnarray}
Q(\theta,\phi) = \frac{1}{\pi}\bra{\theta,\phi}\hat{\rho}^{\nu}_{\mathcal{S}}\ket{\theta,\phi}
\end{eqnarray}
which describes the semiclassical phase space distribution. As shown in Fig.\ref{fig:4_qs}(e-h), the Husimi distribution of such eigenstates exhibit maximum density around the unstable FPs, indicating a localization in phase space. Note that, we have plotted the Husimi distributions in the $z_i=\cos{\theta_i}$ and $\phi_i$ plane, to compare it with the classical phase portraits. Following this prescription, we identify the scars of trivial FPs T$_\pm$, non trivial FPs NT$_\pm$, FP-$\pi$ and 2-cycles TC$_{2\pm}$ shown in Fig.\ref{fig:1_classical_analysis}(a). Because of the complementary behavior of the dynamical classes, we only show the scars corresponding to class II in Fig.\ref{fig:4_qs}. Here we emphasize that, these scars can also be observed in KT model with corresponding dynamical class except the scar of FP-$\pi$. In appendix \ref{quantum_scar_KT}, we discuss in details such scarring phenomena in KT model, since the experimental realization of this model opens up the possibility to detect the scars. 
 
Rather than the entanglement entropy, the statistical analysis of the Floquet eigenstates provides an effective way to distinguish the scarred states. For this purpose, we decompose the Floquet eigenstates $\ket{\phi_{\nu}}=\sum_{i}\phi^{i}_{\nu}\ket{i}$ in the computational basis $\ket{i}$. In the chaotic regime, according to Berry's conjecture \cite{berry_conj}, the eigenstates behave as random states and the probability distribution of their components $\eta = |\phi^{i}_{\nu}|^2\mathcal{N}$, follows the well known Porter-Thomas (PT) distribution ${\rm P}(\eta) = (1/\sqrt{2\pi\eta})\exp(-\eta/2)$ \cite{haake_book}. Consequently, the Shannon entropy $S_{\rm Sh}=-\sum_{i}|\phi^{i}_{\nu}|^2{\rm log}|\phi^{i}_{\nu}|^2$ of such ergodic states attains the value ${\rm log}(0.48\mathcal{N})$  corresponding to its COE limit \cite{Izrailev_2,rigol_alessio}. As the system  approaches to chaos, the Shannon entropy $S_{\rm Sh}$ of the Floquet eigenstates forms a band like structure around the COE limit, however for some eigenstates, $S_{\rm Sh}$ is found to be significantly lower than this limit, which we identify as eigenstates bearing scar, as shown in Fig.\ref{fig:5_NT_+_scar}(a,b). Apart from the scars of the FPs and 2-cycles, we find other type of scars, which resemble the shape of unstable orbits around such FPs, as shown in Fig.\ref{fig:5_NT_+_scar}(e,f). As seen from Fig.\ref{fig:5_NT_+_scar}(c), unlike the ergodic states, the eigenstates bearing scar deviate from the PT distribution, leading to the violation of Berry's conjecture \cite{mondal,sinha1}. However, magnitude of such deviation, as well as the overlap with the corresponding coherent state, depends on the degree of scarring, which decreases with enhanced instability of the underlying dynamics, as shown in Fig.\ref{fig:5_NT_+_scar}(d). Consequently, the scars gradually disappear as the system enters into deep chaotic regime and eventually becomes uniformly ergodic. 

\begin{figure}
\centering
\includegraphics[height=14cm,width=7.0cm]{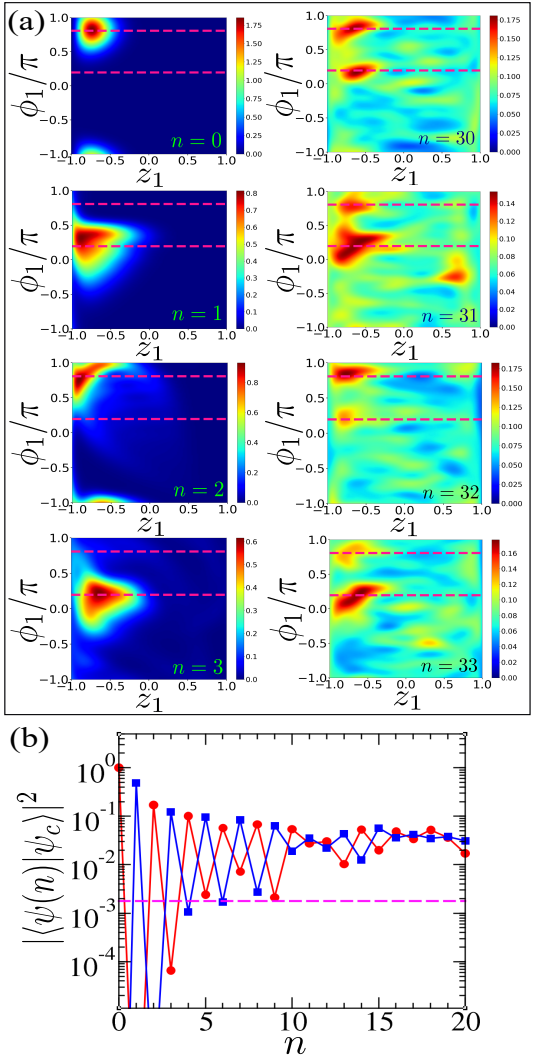}
\caption{Dynamical signature of quantum scar corresponding to a 2-cycle (TC$_{2+}$) of KCT: (a) Stroboscopic time evolution ($t=nT$) of the Husimi distribution exhibiting periodic oscillation of phase space density between the fixed points of unstable TC$_{2+}$ at $\mu=4.34$. The pink dashed lines denote the two fixed points of this 2-cycle represented by $s'$ , $s''$ and we choose the initial coherent state representing one of them. (b) Overlap (survival probability) of the stroboscopically evolved  state $\ket{\psi(n)}$  with the coherent states $\ket{\psi_c}=\ket{s'}$ (red line) and with $\ket{\psi_c}=\ket{s''}$ (blue line), corresponding to two points of 2-cycle. Complementary behavior of both the overlaps reflects periodic oscillation as observed in (b). The pink dashed line in (b) represents the COE limit of survival probability.}
\label{fig:6_scar_2_cycle}
\end{figure}

Unlike the scar of FPs, the quantum scarring of 2-cycles such as TC$_{2\pm}$ has an interesting dynamical feature, since it is the shortest orbit representing the oscillation between two phase space points $s'$ and $s''$. Here, we discuss the dynamical manifestation of the unstable 2-cycles TC$_{2\pm}$. Starting from the initial coherent state representing one of these points of the 2-cycle, we obtain the Husimi distribution of the stroboscopically evolved state $\ket{\psi(n)}$ successively, exhibiting the oscillation of phase space density between these two points. In Fig.\ref{fig:6_scar_2_cycle} (a), such oscillations of Husimi distribution is shown for the 2-cycle TC$_{2+}$. As a result of the instability of this 2-cycle, the Husimi distribution spreads out, however, the quantum scar can still be identified from the accumulation of the phase space density around these points of TC$_{2+}$. We also calculate the overlap of the time evolved state $\ket{\psi(n)}$ with the coherent states $\ket{s'}$ and $\ket{s''}$ corresponding to two fixed points of this 2-cycle. As depicted in Fig.\ref{fig:6_scar_2_cycle}(b), the complementary behavior of the oscillations of the overlaps clearly captures the dynamics between the two points of unstable TC$_{2+}$.

\subsection{Signature of scars from FOTOC dynamics}
\label{FOTOC_scars}
In recent years, a technique known as `out-of-time-order correlator' (OTOC) has been extensively studied to probe quantum many body chaos and scrambling phenomena \cite{stanford2,maldacena,K_hashimoto,Rozenbaum,Garttner1,
butterfly_effect,Swingle1,Fazio_otoc,A_M_rey2,Santos_otoc,Garcia_mata,
sray1,lakshminarayan,sudip_otoc}. The OTOC for two operators $\hat{W}$ and $\hat{V}$ is defined as,
\begin{eqnarray}
O(t) = {\rm Tr}\hat{\rho}_0 \hat{W}^\dagger(t)\hat{V}^\dagger(0)\hat{W}(t)\hat{V}(0)
\end{eqnarray}
where $\hat{W}(t)$ denotes the operator at time $t$ and $\hat{\rho}_0$ is the initial density matrix. For unitary operators $\hat{W}$ and $\hat{V}$, the growth rate of $1-{\rm Re}(O(t))$ can yield the Lyapunov exponent in quantum systems \cite{Rozenbaum,A_M_rey2,Santos_otoc}, moreover its saturation value can provide an alternate measure to quantify the degree of ergodicity \cite{Garcia_mata,sray1,lakshminarayan,sudip_otoc}. For the pure states, the OTOC can be generalized to `Fidelity-OTOC' (FOTOC) $\mathcal{F}_G$ for a hermitian operator $\hat{G}$, which is defined for $\hat{W}=e^{\imath\delta\phi\hat{G}}$ and  $\hat{V}=\hat{\rho}_0=\ket{\psi(0)}\bra{\psi(0)}$ corresponding to the initial state $\ket{\psi(0)}$ \cite{A_M_rey2,Santos_otoc}. In the limit $\delta \phi \ll 1$, the FOTOC can be written in terms of the fluctuation $f_G$ of the corresponding operator $\hat{G}$,  
\begin{eqnarray}
1-\mathcal{F}_{G} \approx \delta\phi^2 \left(\langle \hat{G}^2\rangle - \langle \hat{G}\rangle^2\right) \equiv  \delta\phi^2 f_{G}.
\end{eqnarray}
which simplifies the computation of $\mathcal{F}_G$ and makes it suitable for collective systems. In the perturbative regime ($\delta\phi\ll 1$), the dynamics of FOTOC, as well the growth rate of $1-\mathcal{F}_G$ can be captured from the time evolution of the corresponding fluctuation $f_{G}$, which can successfully capture the instability exponent \cite{Santos_otoc} and scrambling \cite{A_M_rey2} in quantum system.
\begin{figure}[H]
\centering
\includegraphics[height=7cm,width=8.7cm]{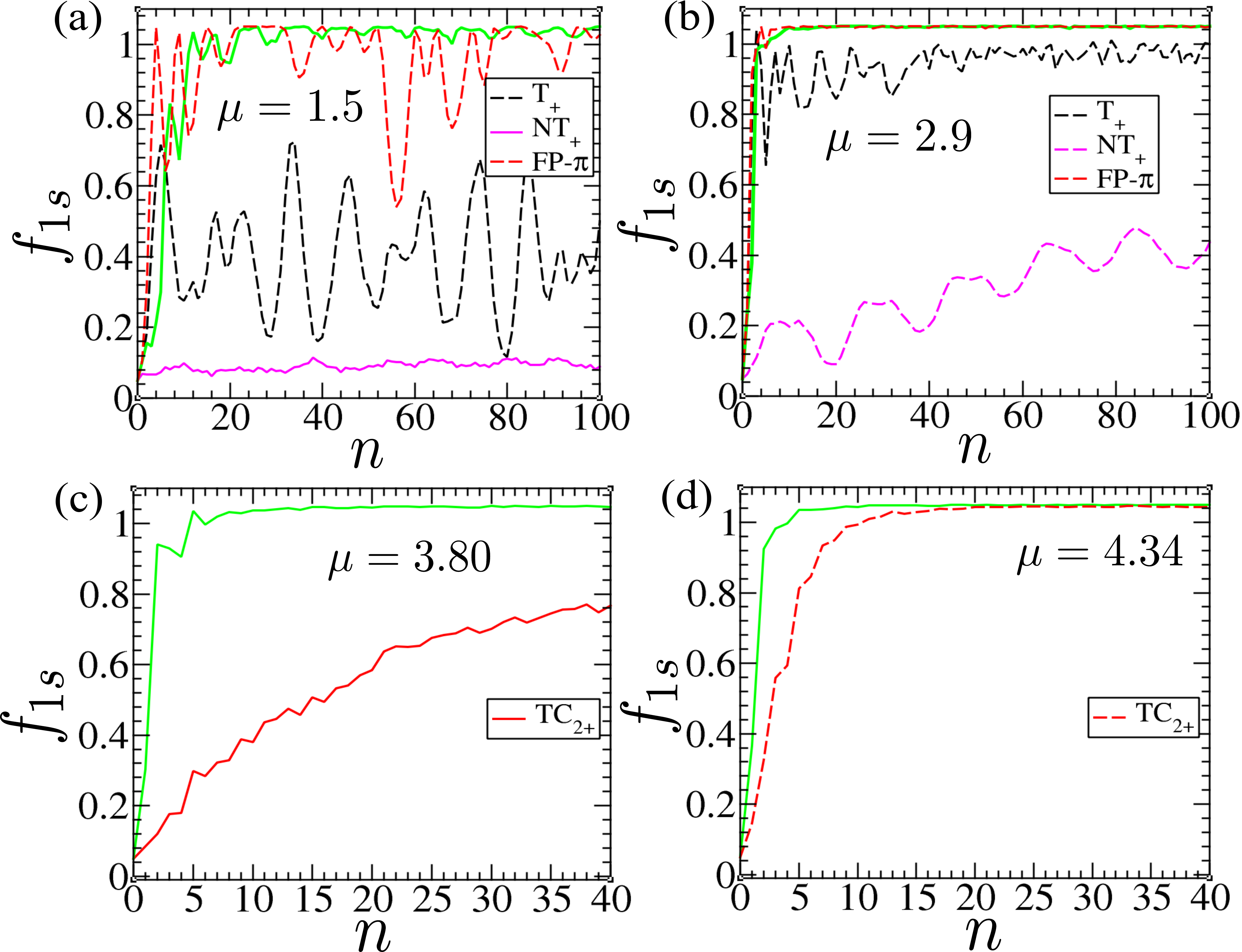}
\caption{(a-b) Comparison of FOTOC dynamics starting from initial coherent states corresponding to different stable (solid) and unstable (dashed) FPs for two different values of $\mu$. As NT$_+$ becomes unstable, the effect of scarring is reflected from the growth and oscillation of FOTOC  (shown in (b)), which is contrasted with its behavior when NT$_+$ is stable (shown in (a)). (c-d) Dynamics of FOTOC starting from the initial state representing one of the fixed points of TC$_{2+}$ (red line) for two different values of $\mu$ corresponding to stable (solid) and unstable (dashed) TC$_{2+}$. In all the cases, the green line denotes FOTOC dynamics for a random initial coherent state belonging to the chaotic region.}
\label{fig:7_FOTOC_dynamics}
\end{figure}
The dynamics of $1-\mathcal{F}_G$ for collective spin systems like KCT can alternatively be studied from the fluctuation $f_G$ for suitable spin operators with $\hat{G}=\hat{S}_{ia}/S$. To detect the dynamical signature of quantum scars, we investigate the dynamics of total fluctuation $f_{is}$ of all the components of a particular spin sector, which is given by,
\begin{eqnarray}
f_{is} = \sum_{a=x,y,z}f_{isa}=\sum_{a=x,y,z} (\langle \hat{S}^2_{ia}\rangle - \langle \hat{S}_{ia} \rangle^2)/S^2.
\end{eqnarray}
 
In the mixed phase space region, we study the dynamics of total spin fluctuations $f_{is}$ starting from the initial coherent states representing the stable (unstable) FPs and 2-cycles. For stable FPs surrounded by the regular regions of phase space, the $f_{is}$ exhibits oscillatory behavior with very small amplitude, whereas for initial coherent state belonging to the chaotic region, $f_{is}$ grows rapidly and saturates to unity. On the other hand, the $f_{is}$ for unstable FPs exhibits an intermediate behavior with slower growth rate and large oscillations (see Fig.\ref{fig:7_FOTOC_dynamics}(a,b)) indicating the scarring phenomena. Such behavior can be contrasted with that of stable FPs, which can have relevance in experiments to distinguish quantum scars from unstable FPs in the mixed phase space region. Similarly, we also study the dynamics of $f_{is}$ for the stable and unstable 2-cycle TC$_{2+}$ shown in  Fig.\ref{fig:7_FOTOC_dynamics} (c,d), which exhibits larger growth rate as the 2-cycle becomes unstable. The reduction of degree of scarring due to enhanced dynamical instability can also be captured from FOTOC dynamics, since both the growth rate and saturation corresponding to a scarred state increases as the system approaches to a more chaotic regime with increasing kicking strength $\mu$.

\section{Conclusion}
\label{CONCLUSION} 
In the present work, we investigated the local ergodic behavior of a coupled top model subjected to periodic kicking and unveil its connection with the underlying phase space dynamics, which plays a crucial role in the formation of quantum scars. With increasing the kicking strength, the system undergoes a crossover from regular to chaotic dynamics. In the mixed phase space, the regular regions around the stable fixed points (FPs) and 2-cycles give rise to strong deviation of entanglement entropy and survival probability from their ergodic limit, revealing the local ergodic behavior. As the unstable FPs and 2-cycles disappear from the phase portrait, their reminiscence can still be visible through deviation from the ergodic limit, exhibiting quantum scarring phenomena. Also, we  discuss the methods for identification of scars in Floquet eigenstates from their statistical properties and Shannon entropy. Such eigenstates carrying the scars exhibit violation of Berry's conjecture in contrast to the ergodic states. However, even after instability, the trajectories remain localized near such unstable FPs, which essentially gives rise to phase space localization in scarred states, as visible in Husimi distribution. Apart from the FPs, we have also identified the scars of 2-cycles, giving rise to oscillation between two phase space points.

We have shown how quantum scars in mixed phase space can be distinguished from both the stable FPs and ergodic states, by the FOTOC dynamics, which can serve as an efficient method for its experimental detection. The implementation of FOTOC has already been done in trapped ion simulators \cite{A_M_rey2}, which can also serve as a platform to engineer collective spin models \cite{otoc_exp_2}. The experimental realization of kicked top model in cold atom setup \cite{KT_exp}, and in superconducting qubits \cite{Neill} has opened up the immediate possibility to investigate the quantum scarring phenomena.
 
\section*{ACKNOWLEDGMENT}
 We thank Hans Kroha and Sayak Ray for comments and discussion.

\appendix
\setcounter{figure}{0}
\section{Derivation of stroboscopic evolution of spin operators} 
\label{derivation_dynamical_map}
In the Heisenberg picture, the stroboscopic time evolution of the spin operators can be written in terms of $\hat{\mathcal{F}}$ as, $\hat{S}_{ia}^{(n+1)}=\hat{\mathcal{F}}^{\dagger n+1}\hat{S}_{ia}\hat{\mathcal{F}}^{n+1}=\hat{\mathcal{F}}^{\dagger}\hat{S}^{(n)}_{ia}\hat{\mathcal{F}}$, where $i=1,2$; $a=x,y,z$ and $\hat{S}_{ia}^{(n)}$ denotes the operator at time $t=nT$. Following this prescription, here we only derive the equation of motion for the $z$ component of spin $\hat{S}_{1z}$,
\begin{eqnarray}
\hat{S}_{1z}^{(n+1)} &=& \hat{\mathcal{F}}^{\dagger n}(\hat{\mathcal{F}}^\dagger \hat{S}_{1z}\hat{\mathcal{F}})\hat{\mathcal{F}}^{n}\notag \\
 &=&\hat{\mathcal{F}}^{\dagger n}(e^{-\imath (\hat{S}_{1x}+\hat{S}_{2x})T}e^{-\imath\frac{\mu}{S}\hat{S}_{1z}\hat{S}_{2z}} \hat{S}_{1z} \notag\\
 &&e^{\imath\frac{\mu}{S}\hat{S}_{1z}\hat{S}_{2z}}e^{\imath (\hat{S}_{1x}+\hat{S}_{2x})T} )\hat{\mathcal{F}}^{n} \notag \\
 &=&\hat{\mathcal{F}}^{\dagger n}(e^{-\imath (\hat{S}_{1x}+\hat{S}_{2x})T} \hat{S}_{1z} e^{\imath (\hat{S}_{1x}+\hat{S}_{2x})T} )\hat{\mathcal{F}}^{n} \notag \\
 &=&\hat{\mathcal{F}}^{\dagger n}(\hat{S}_{1z}\cos{T}-\hat{S}_{1y}\sin{T})\hat{\mathcal{F}}^{n}\notag\\
 &=&\hat{S}^{(n)}_{1z}\cos{T}-\hat{S}^{(n)}_{1y}\sin{T}
\end{eqnarray}
where we have used the commutation relation $[\hat{S}_{ia},\hat{S}_{jb}]=\imath\epsilon_{abc}\delta_{ij}\hat{S}_{ic}$ and the following operator identity, 
\begin{eqnarray}
e^{t\hat{X}}\hat{Y}e^{-t\hat{X}} = \hat{Y} + t[\hat{X},\hat{Y}] +\frac{t^2}{2}[\hat{X},[\hat{X},\hat{Y}]] + ...
\end{eqnarray}
In similar manner, the equations of motion for other components can be derived. To obtain the classical map (see Eq.\eqref{classical_map} of Sec.\ \ref{CLASSICAL_ANALYSIS}), we have redefined the operators $\hat{S}_{ia}$ as $\hat{s}_{ia} = \hat{S}_{ia}/S$, which can be treated as classical variables in the limit $S\rightarrow\infty$, since the commutation relation [$\hat{s}_{ia},\hat{s}_{jb}$] vanishes as $1/S$. 

\setcounter{figure}{0}
\begin{figure}[H]
\centering
\includegraphics[height=4.2cm,width=8.7cm]{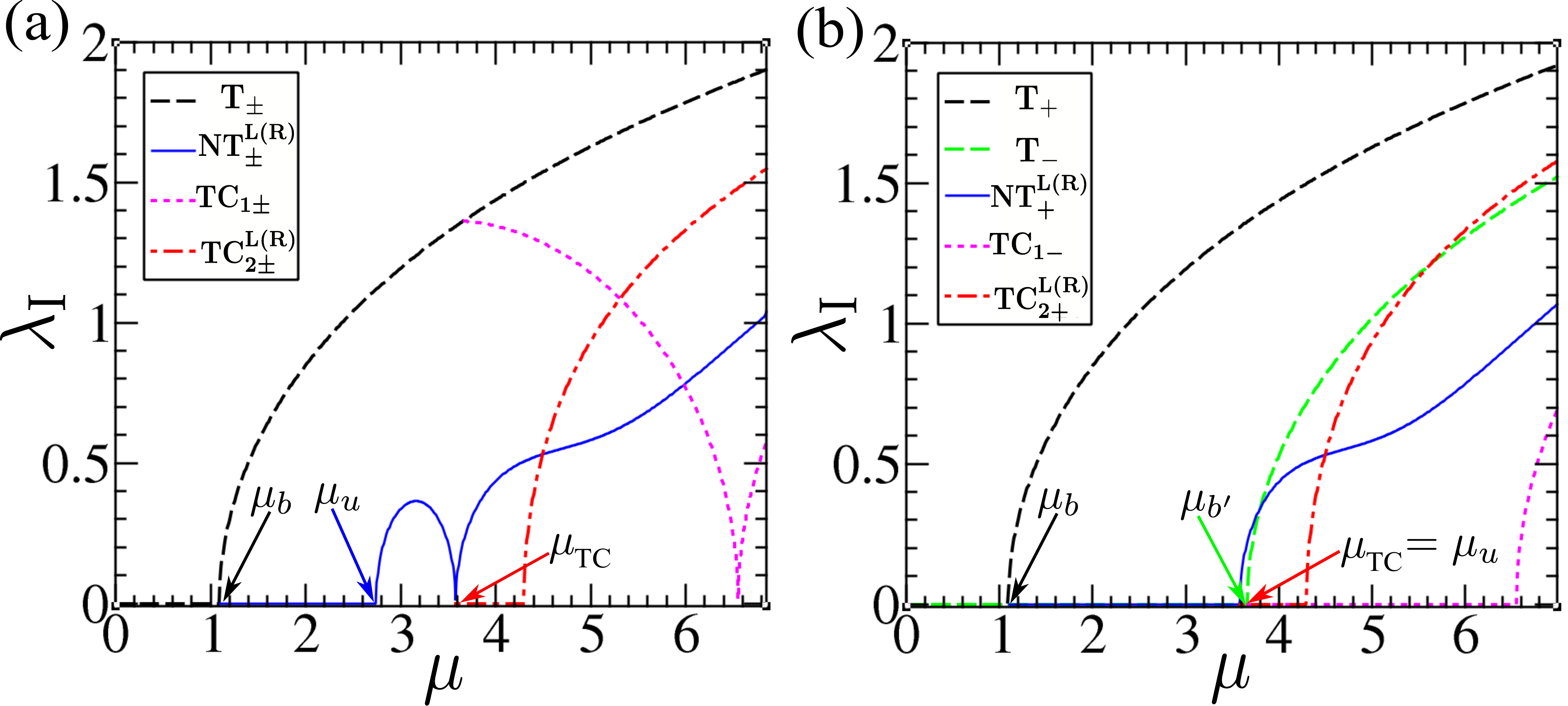}
\renewcommand{\thefigure}{B\arabic{figure}}
\caption{Dynamical instability of different FPs and 2-cycles (as mentioned in the figure). The instability exponent $\lambda_{\rm I}$ with increasing kicking strength $\mu$  for (a) KCT and (b) ferromagnetic KT model corresponding to dynamical class II.}
\label{fig:1_app}
\end{figure}
\begin{figure*}
	\centering
	\includegraphics[height=8.2cm,width=18.3cm]{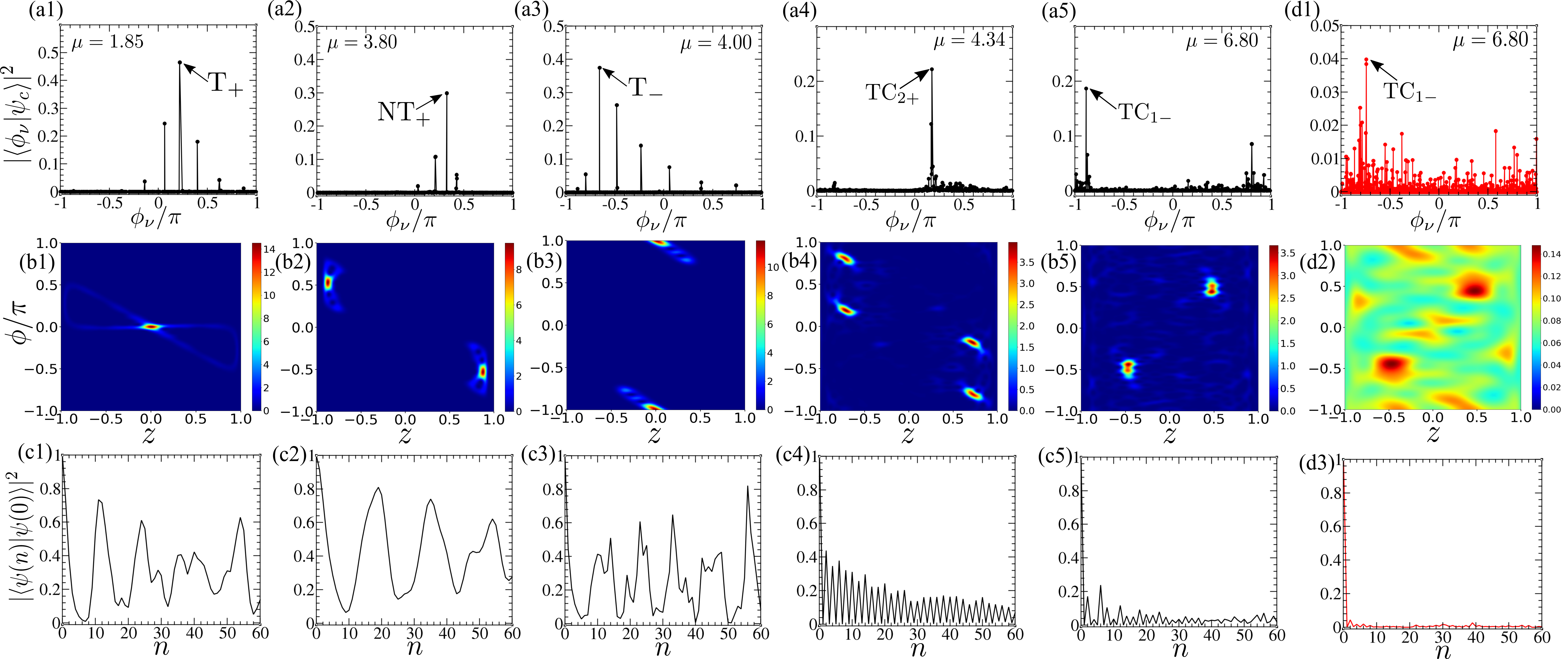}
	\renewcommand{\thefigure}{C\arabic{figure}}
	\caption{Identification of different scars in ferromagnetic KT model and their dynamical signature. (a1-a5) Overlap of Floquet eigenstates $\ket{\phi_{\nu}}$ with coherent state representing different FPs and 2-cycles of KT model. (b1-b5) Husimi distribution of the eigenstate with maximum overlap (marked by arrowheads in (a1-a5)) revealing the quantum scars of corresponding unstable FPs and 2-cycles. (c1-c5) Survival probability $|\langle \psi(n)|\psi(0)\rangle|^2$ where $\ket{\psi(0)}$ represents the initial coherent state of the above mentioned unstable FPs exhibiting revival phenomena due to scarring. The last column (d1-d3) shows the scarring of TC$_{1-}$ in KCT, where the same quantities are compared with that of the ferromagnetic KT model shown in column (a5-c5).}
	\label{fig:2_app}
\end{figure*}

\section{Stability analysis}
\label{stability_analysis}
The stability of the fixed points (FPs) and 2-cycles can be analyzed by linearizing the classical map given in Eq.\eqref{classical_map} for small fluctuation around them. Following the standard procedure in \cite{lichtenberg,strogatz}, we construct the Jacobian matrix ${\rm J}$, whose matrix elements are given by ${\rm J}_{\alpha \beta}=\partial s^{(n+1)}_\alpha/\partial s^{(n)}_\beta$, where $n$ is the stroboscopic time and  $\alpha,\beta= 1,2...,6$ are the indices of the array $s = \{s_{1x},s_{1y},s_{1z},s_{2x},s_{2y},s_{2z}\}$ representing the phase space point of the two spin system. We calculate the instability of an unstable FP represented by $s^{*}$ from the instability exponent $\lambda_I=\ln (|j_m|)>0$, where $j_m$ is the eigenvalue of the Jacobian matrix J($s^{*}$) with maximum magnitude, evaluated at $s^{*}$. Similarly, the stability of a 2-cycle can be obtained from the matrix $\tilde{\rm J} = {\rm J}(s'){\rm J}(s'')$, where the Jacobian matrices ${\rm J}$ are evaluated at the corresponding fixed points $s'$ and $s''$ of the 2-cycle. The corresponding instability exponent of the 2-cycle is given by $\lambda_\text{I}=(1/2)\ln (|\tilde{j}_m|)>0$, where $\tilde{j}_m$ is the eigenvalue of matrix ${\tilde{\rm J}}$ with maximum magnitude. The stability of FP (2-cycles) is ensured if the magnitude of all eigenvalues of ${\rm J}$ ($\tilde{\rm J}$) are unity \cite{lichtenberg}. In the present case, as a result of the constraint, $s^2_{ix}+s^2_{iy}+s^2_{iz}=1$ (for $i=1,2$), the magnitude of two eigenvalues of ${\rm J}$ and $\tilde{\rm J}$ always remain unity. We compute the instability exponents $\lambda_\text{I}$ of the FPs and 2-cycles of KCT, that we discussed in Sec.\ref{CLASSICAL_ANALYSIS}, with increasing kicking strength $\mu$ (see Fig.\ref{fig:1_app}(a)). We also compare them with the instability of the FPs and 2-cycles of ferromagnetic KT model (corresponding to dynamical class II), shown in Fig.\ref{fig:1_app}(b). Here it is important to note that the 2-cycle TC$_{1\pm}$ is present in KCT but remains always unstable.

\begin{figure}
\centering
\includegraphics[height=4cm,width=9cm]{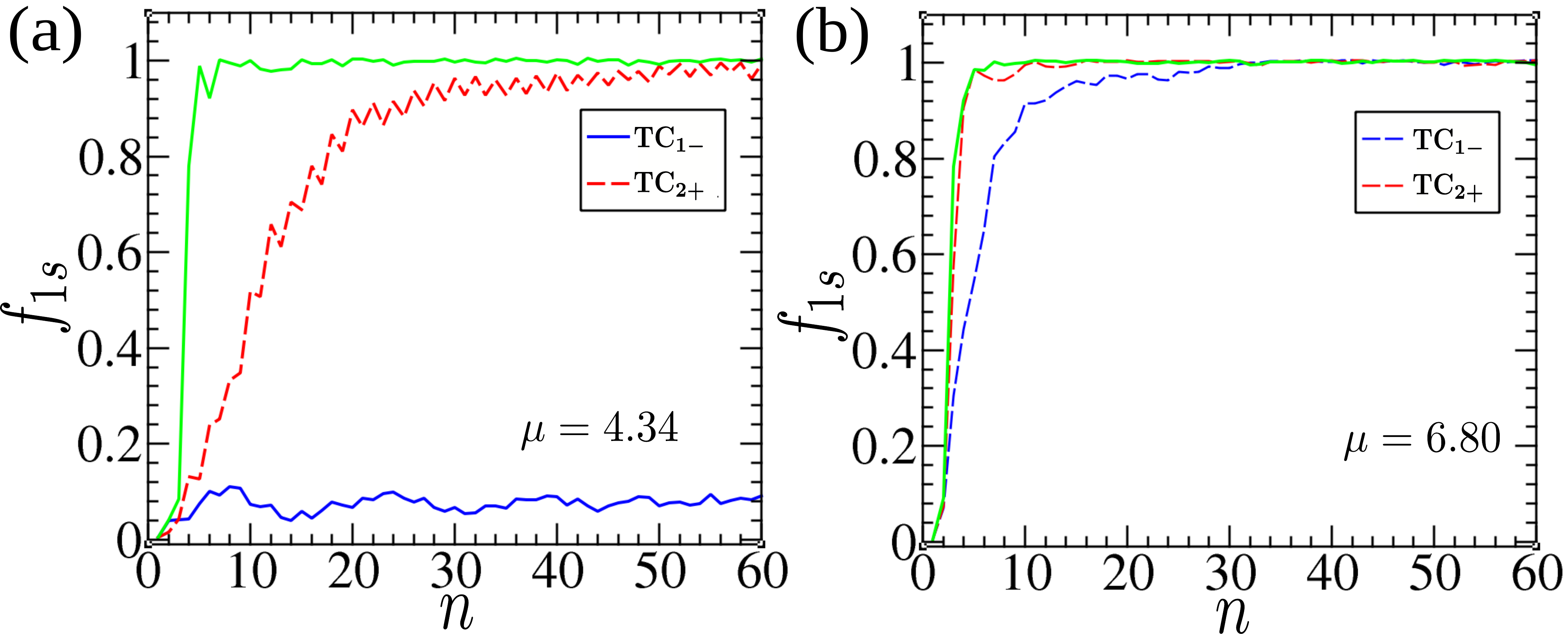}
\renewcommand{\thefigure}{C\arabic{figure}}
\caption{Signature of scars of 2-cycles in ferromagnetic KT model from FOTOC dynamics: (a-b) Comparison of FOTOC dynamics starting from initial coherent states corresponding to 2-cycles TC$_{2+}$ (red line) and TC$_{1-}$ (blue line) for two different values of $\mu$. The stable (unstable) 2-cycles are shown by solid (dashed) line. In both the figures, the green line represents the same for the initial coherent state belonging to the chaotic regime. Scarring of TC$_{1-}$ is captured from the larger growth rate of FOTOC  as it becomes unstable (shown in (b)), which can be contrasted to its behavior when  it is stable (shown in (a)). With increasing $\mu$, the FOTOC for unstable TC$_{2+}$ (shown in (b)) becomes almost similar to that of an ergodic state showing the reduction of degree of scarring.}
\label{fig:3_app}
\end{figure}

\section{quantum scars in ferromagnetic kicked top model}
\label{quantum_scar_KT}
As shown in subsection \ref{DYNAMICAL_CLASSES}, the dynamics of kicked coupled top (KCT) can be divided into two classes (I)II corresponding to (anti)ferromagnetic kicked top (KT) model. Similar scarring phenomena can also be observed in the KT model, which we discuss in this appendix. Here we analyze the scars of unstable FPs and 2-cycles corresponding to the ferromagnetic KT model, which is shown in Fig.\ref{fig:2_app}. The scarred eigenstates $\ket{\phi_{\nu}}$ are identified from the large overlap with the coherent states $\ket{\psi_c}$ representing the unstable FPs and 2-cycles (see Fig.\ref{fig:2_app}(a1-a5)). The scars in such eigenstates can also be visualized from the Husimi distribution localized around those FPs and 2-cycles, as depicted in Fig.\ref{fig:2_app}(b1-b5). Here we have identified scars of trivial FPs T$_\pm$, non trivial FPs NT$_+$ and the 2-cycles TC$_{2+}$, TC$_{1-}$ (see Fig.\ref{fig:1_classical_analysis}(c) for the fixed points of ferromagnetic KT model), which also manifest revivals in the corresponding survival probabilities $F(n)$ (see Fig.\ref{fig:2_app}(c1-c5)). We also emphasize, although the 2-cycle TC$_{1-}$ is present in both the ferromagnetic KT as well as KCT model, the degree of scarring in KCT is weaker due to larger instability generated because of mixing between the two dynamical classes (class I and II), as reflected from the comparison of Husimi distributions shown in Fig.\ref{fig:2_app}(b5) and \ref{fig:2_app}(d2). Such scars can also be identified from FOTOC dynamics, which is shown for 2-cycles TC$_{1-}$ and TC$_{2+}$. As shown from the comparison between Fig.\ref{fig:3_app}(a,b), it is evident that the onset of dynamical instability of TC$_{1-}$ leads to a rapid enhancement in growth and magnitude of FOTOC, which is however slower than an ergodic state. As a result of enhanced instability, the degree of scarring of TC$_{2+}$ reduces and corresponding FOTOC becomes almost indistinguishable from that of an ergodic state, which is shown in Fig.\ref{fig:3_app}(b). Here we point out that the KT model has already been realized experimentally in cold atom systems \cite{KT_exp}, as well in superconducting qubits \cite{Neill}, which opens up an immediate possibility to investigate such quantum scarring phenomena, particularly the scar of 2-cycles can also be diagnosed from the FOTOC dynamics.

\end{document}